\documentclass[conference]{IEEEtran}
\usepackage{cite}
\usepackage{amsmath,amssymb,amsfonts}
\usepackage{algorithmic}
\usepackage{graphicx}
\usepackage{textcomp}
\usepackage{xcolor}
\usepackage{multirow}
\usepackage{graphicx}
\usepackage{booktabs}
\usepackage{url}
\usepackage{float}
\usepackage[caption=false,font=footnotesize]{subfig}

\def\BibTeX{{\rm B\kern-.05em{\sc i\kern-.025em b}\kern-.08em
    T\kern-.1667em\lower.7ex\hbox{E}\kern-.125emX}}
\begin{document}

\title{An Effective Strategy for Modeling Score Ordinality and Non-uniform Intervals in Automated Speaking Assessment  \\

}

\author{
    \IEEEauthorblockN{
    Tien-Hong Lo\IEEEauthorrefmark{1},
    Szu-Yu Chen\IEEEauthorrefmark{2},
    Yao-Ting Sung\IEEEauthorrefmark{2}, and
    Berlin Chen\IEEEauthorrefmark{1}
    }
    \IEEEauthorblockA{
    \IEEEauthorrefmark{1} 
    Department of Computer Science and Information Engineering, National Taiwan Normal University
    }
    \IEEEauthorblockA{
    \IEEEauthorrefmark{2}
    Department of Educational Psychology and Counseling, National Taiwan Normal University
    }
}
\maketitle

\begin{abstract}
A recent line of research on automated speaking assessment (ASA) has benefited from self-supervised learning (SSL) representations, which capture rich acoustic and linguistic patterns in non-native speech without underlying assumptions of feature curation. However, speech-based SSL models capture acoustic-related traits but overlook linguistic content, while text-based SSL models rely on ASR output and fail to encode prosodic nuances. Moreover, most prior arts treat proficiency levels as nominal classes, ignoring their ordinal structure and non-uniform intervals between proficiency labels. To address these limitations, we propose an effective ASA approach combining SSL with handcrafted indicator features via a novel modeling paradigm. We further introduce a multi-margin ordinal loss that jointly models both the score ordinality and non-uniform intervals of proficiency labels. Extensive experiments on the TEEMI corpus show that our method consistently outperforms strong baselines and generalizes well to unseen prompts.
\end{abstract}

\begin{IEEEkeywords}
automated speaking assessment, self-supervised learning, multi-aspect features, ordinal classification, non-uniform score interval.
\end{IEEEkeywords}

\section{Introduction}

With the rapid advances in computing technology and the growing global population of second-language (L2) learners, automated speaking assessment (ASA) has garnered considerable attention and plays an increasingly prominent role in computer-assisted language learning (CALL). ASA systems are developed to deliver timely feedback on learners' speaking performance, facilitating autonomous and low-stress improvement in spoken language proficiency. In addition,  ASA systems help reduce the burden on language instructors while offering more consistent and objective evaluations of L2 learners’ speaking proficiency. In light of these technological developments, ASA systems have been widely adopted in recent years to enhance L2 language acquisition within a wide spectrum of CALL use cases \cite{IntroAsaVan2016_hsla}.

\begin{figure}[htbp]
  \centering
  \includegraphics[width=0.6\linewidth]{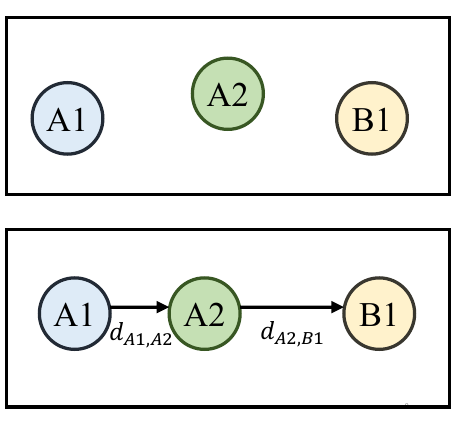}
  \caption{A conceptual illustration depicts the distinction between the nominal and ordinal structure of CEFR proficiency levels. The upper panel treats levels (e.g., A1, A2, B1) as nominal categories, without reflecting any inherent order or distance between them. The lower panel models the ordinal structure by representing directed transitions between proficiency levels (i.e., A1~$\rightarrow$~A2~$\rightarrow$~B1), with distances $d_{a_1,a_2}$ and $d_{a_2,b_1}$ indicating non-uniform score intervals.}
  \label{fig:ordinal_description}
\end{figure}

Early ASA methodologies primarily employed shallow classifiers alongside handcrafted features capturing distinct aspects of language proficiency, including content (e.g., appropriateness and relevance), delivery (e.g., fluency and intonation), and language use (e.g., vocabulary and grammar) \cite{ApaCucchiarini1998_icslp,IntroAsaAnaStrik1999_book,IntroAsaAnasChen2010_slt,IntroAsaAnasBhat2015_spcomm,AsaMoore2015_springer,ApaCoutinho2016_lrec,AsaMuller2009_slate,AesWu2022_rocling,SpeechraterChen2018_ets}. More recently, the emergence of modeling paradigms for self-supervised learning (SSL), such as BERT and its derivatives \cite{BertDevlin2019_naacl}, has brought new opportunities for ASA through the provision of contextualized embeddings. These representations have been effectively leveraged in a variety of assessment tasks, including sentence-level evaluation \cite{AesProtoBertArase2022_emnlp}, essay scoring \cite{AesNadeem2019_bea,AesWu2023_slate}, the evaluation of spoken monologues \cite{AsaMoore2015_springer,AsaCraighead2020_acl}, and among others \cite{Asali2023_slate}. In parallel, the drastic development of speech-based SSL models has further fostered the effectiveness of ASA systems by offering rich acoustic representations that support more sophisticated modeling capabilities \cite{AsaBanno2022_slt,AsaMcknight2023_slate,AsaBanno2023_slate,AsaLo2024_naacl}.

Despite the continued efforts, existing SSL-based ASA approaches remain constrained by modality-specific limitations. Speech-based SSL encoders are adept at modeling acoustic characteristics but often fail to capture the semantic content of learner responses. In contrast, text-based SSL models rely on automatic speech recognition (ASR) outputs, rendering them susceptible to transcription errors and incapable of representing prosodic cues essential for evaluating the delivery aspect. Furthermore, both modalities tend to lack interpretability and overlook explicit indicators such as pitch variation and word-level accuracy. These limitations underscore the necessity of a multi-aspect modeling strategy that integrates handcrafted features with SSL-derived embeddings to exploit the complementary strengths of different modalities \cite{AsaQian2019_icassp,AsaPark2023_asru,AsaPeng2023_slt}. Beyond modality-related constraints, ASA systems must also address challenges posed by the structure of proficiency levels (labels) of international language standards, such as Common European Framework of Reference for Languages (CEFR). Specifically, CEFR levels exhibit an inherent ordinal relationship (e.g., A1 $<$ A2 $<$ B1 $<$ B2), yet many existing methods treat them as nominal categories, disregarding their ordered nature. This modeling simplification may lead to suboptimal training objectives and reduced calibration performance. Moreover, the score intervals between CEFR levels are non-uniform (e.g., the progression from B1 to B2 is not equivalent to that from A1 to A2) \cite{CefrNuiHheilman2008_bea,CefrLord2012_applications}. As illustrated in Figure \ref{fig:ordinal_description}, most prior studies treat all level gaps as equidistant, as is often done in regression or classification settings, falling short of reflecting the underlying learning progression which would hinder interpretability.

To tackle the aforementioned challenges, we explore an innovative ASA modeling approach that integrates handcrafted features with self-supervised embeddings to jointly capture the acoustic and semantic properties of learner speech. In addition, we put forward a multi-margin ordinal (MMO) loss function that explicitly models the ordinal structure and non-uniform intervals inherent in CEFR proficiency levels. Experiments conducted on the TEEMI dataset demonstrate that the proposed framework consistently outperforms strong baselines, particularly in identifying underrepresented proficiency levels such as Pre-A1 and B2. Furthermore, the model exhibits robust generalization to unseen prompts and speakers, highlighting its applicability in real-world ASA scenarios. The primary contributions of this work are at least two-fold:
\begin{enumerate}
\item We propose a multi-aspect ASA framework that integrates handcrafted features with self-supervised embeddings to capture modality-specific characteristics, while addressing the ordinal and non-uniform properties of the CEFR proficiency scale. A series of experiments on the TEEMI dataset reveal substantial improvements in macro-averaged F1 scores over competitive baselines.
\item We present a novel MMO loss function that, to the best of our knowledge, is the first to jointly model the ordinal structure and non-uniform level intervals of CEFR-aligned scores in a fully data-driven manner. This design promotes both the performance and interpretability of the predicted proficiency levels.
\end{enumerate}

\section{Related Work}
Automated speaking assessment (ASA) seeks to evaluate the oral proficiency of second-language (L2) learners, typically through either holistic scores that reflect overall proficiency or analytic scores that assess specific aspects of performance.

\subsection{Handcrafted Features}
Early ASA systems primarily operated with shallow classifiers trained on handcrafted features designed to capture salient aspects of spoken language, including pronunciation, fluency, prosody, and grammar, to name a few \cite{ApaCucchiarini1998_icslp,IntroAsaAnaStrik1999_book,IntroAsaAnasChen2010_slt,IntroAsaAnasBhat2015_spcomm,AsaMoore2015_springer,ApaCoutinho2016_lrec,AsaMuller2009_slate,SpeechraterChen2018_ets}. For example, \cite{IntroAsaAnasChen2010_slt} employed vowel space metrics to evaluate articulatory precision, while \cite{IntroAsaAnasBhat2015_spcomm} incorporated syntactic complexity indicators derived from part-of-speech distributions. Prosodic and rhythmic patterns were examined in \cite{ApaCoutinho2016_lrec}, and syntactic parsing accuracy was investigated in \cite{AsaMoore2015_springer}. Although interpretable, the extraction of handcrafted features often builds on task-specific assumptions and may struggle to generalize across unseen prompts or different task configurations \cite{AsaMuller2009_slate}.

\subsection{Text-based Self-Supervised Features}
The advent of self-supervised learning (SSL) has led to the widespread adoption of contextualized textual embeddings in ASA. Models such as BERT \cite{BertDevlin2019_naacl} have achieved strong performance across various assessment tasks, including, among others, essay scoring \cite{AesNadeem2019_bea,AesWu2023_slate}, readability prediction \cite{AesProtoBertArase2022_emnlp}, and spoken dialogue evaluation \cite{AsaCraighead2020_acl,Asali2023_slate}. These models effectively capture semantic and syntactic features; however, they depend on ASR-generated transcripts, which are susceptible to recognition errors and incapable of preserving prosodic and phonetic information vital for assessing delivery quality.

\subsection{Speech-based Self-Supervised Features}
Speech-based SSL models, such as wav2vec 2.0 \cite{SslLarochelle2020_nips}, facilitate direct modeling of raw acoustic signals and are capable of encoding fine-grained phonetic and prosodic representations without recourse to ASR transcriptions. Prior work has demonstrated the utility of such models for CEFR-level classification \cite{AsaBanno2022_slt,AsaBanno2023_slate} and modality comparison. \cite{AsaMcknight2023_slate} extended these models to conversational contexts, while \cite{AsaLo2024_naacl} combined prototypical embeddings with loss re-weighting strategies to mitigate issues related to label imbalance. Although effective in modeling delivery-related features, speech-based SSL models often lack the semantic richness which is arguably necessary for evaluating content.
\subsection{Multi-aspect Features}
To capture the multifaceted nature of spoken performance, recent efforts have focused on multi-aspect modeling regimes that jointly predict scores for aspects such as delivery, content, and language use using parallel or hierarchical architectures \cite{ApaChao2022_apsipa,ApaHe2024_apsipa,ApaChao2023_interspeech}. While most of the prior studies center on analytic scoring, some have extended multi-aspect modeling to holistic speaking assessment. For example, \cite{AsaQian2019_icassp} incorporated multi-aspect features into holistic scoring, and \cite{AsaPeng2023_slt} introduced soft-label modeling in the context of speaking assessment. \cite{AsaPark2023_asru} systematically compared wav2vec 2.0 and BERT for different scoring aspects, revealing that the former excels in modeling delivery, while the latter shows slight superiority for content-related tasks. A fusion of both modalities was found to yield the best overall performance.

\begin{table}[!t]
\centering
\caption{Number of speakers for each CEFR proficiency level in the TEEMI dataset.}
\label{tab:teemi_dataset_split}
\resizebox{\columnwidth}{!}{%
\begin{tabular}{lc|ccclllll}
\hline
\textbf{Task} & \textbf{Usage} & \textbf{Pre-A} & \textbf{A1} & \textbf{A1+} & \textbf{A2} & \textbf{A2+} & \textbf{B1} & \textbf{B1+} & \textbf{B2} \\ \hline
\multirow{3}{*}{A01} & Train      & 34 & 61  & 76  & 156 & 150 & 169 & 79  & 65  \\
                     & Valid      & 8  & 16  & 19  & 38  & 39  & 43  & 23  & 12  \\
                     & Test       & 11 & 20  & 23  & 49  & 50  & 48  & 32  & 15  \\ \hline
A02                  & Unseen     & 9  & 7   & 12  & 19  & 12  & 26  & 23  & 15  \\
B02                  & Unseen     & 15 & 14  & 21  & 41  & 48  & 62  & 31  & 16  \\
C01                  & Unseen     & 10 & 12  & 9   & 17  & 16  & 21  & 18  & 16  \\ \hline
Total                & \textbf{-} & 87 & 130 & 160 & 320 & 315 & 369 & 206 & 139 \\ \hline
\end{tabular}%
}
\end{table}

\begin{figure}[!t]
  \centering
  \includegraphics[width=0.9\linewidth]{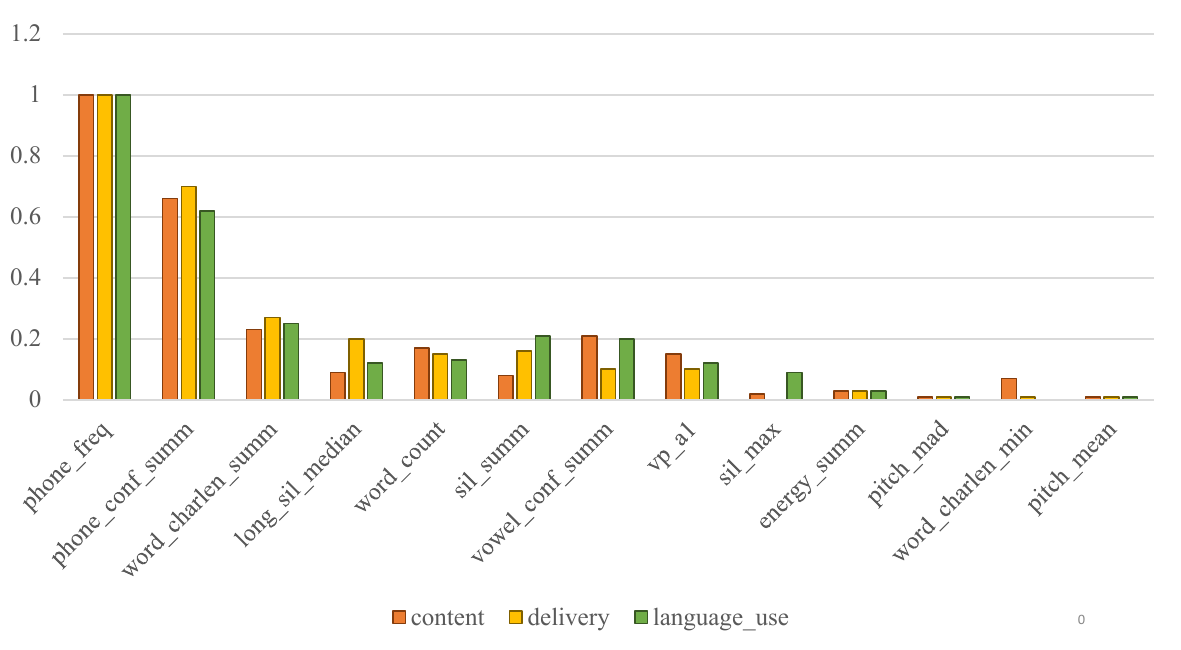}
  \caption{Relative importance of handcrafted features for content, delivery, and language use prediction on the A01 test set from the TEEMI corpus, computed using random forest regressors trained separately for each scoring aspect.}
  \label{fig:feature_importance}
\end{figure}

\section{Dataset}

To evaluate the effectiveness of the proposed ASA framework, this study employs the \textbf{TE}st for \textbf{E}nglish-\textbf{M}edium \textbf{I}nstruction (TEEMI) corpus \cite{CorpusChen2024_teemi_interspeech}, a proprietary dataset curated for research on English-medium instruction (EMI) and ASA. The corpus consists of spontaneous spoken English responses produced by L2 learners at the undergraduate and graduate levels. Each spoken response of an L2 learner is annotated with one holistic score and three analytic scores (i.e., content, language use, and delivery) based on CEFR-aligned rubrics. The process of annotation was carried out by at least three trained raters per response, and the final label was determined by majority voting to ensure reliability.

The component of the TEEMI corpus includes three task formats: general listen and answer (A), situational question and answer (B), and thematic question and answer (C). In this study, we focus on a subset consisting of tasks A01, A02, B02, and C01, yielding a total of 8,214 responses. Model training and validation are performed solely on A01, which contains 6,152 responses from 1,231 speakers. The remaining tasks (A02, B02, and C01) are held out to evaluate the  ability of ASA models to generalize to previously unseen prompts. The detailed CEFR-level distributions across each task and the corresponding partitions are illustrated in Table~\ref{tab:teemi_dataset_split}.

On a separate front, to testify whether tradition handcrafted features have good or poor generalization capabilities across prompts and task types, we analyze the importance of each feature by training independent random forest regressors for content, delivery, and language use, following the approach and feature definitions provided in \cite{AesWu2022_rocling} and \cite{SpeechraterChen2018_ets}. As shown in Figure~\ref{fig:feature_importance}, the efficacy of individual features varies across scoring aspects. For instance, phoneme frequency statistics (\texttt{phone\_freq}) and total silence duration (\texttt{sil\_summ}) are highly informative for delivery scoring, whereas the frequency of CEFR-A1-level vocabulary items (\texttt{vp\_a1}) and word count (\texttt{word\_count}) are more indicative of content assessment. This observation motivates the incorporation of SSL representations, which are expected to capture more comprehensive information for ASA.

% \begin{figure*}[htbp]
%     \centering
%     \includegraphics[width=0.85\linewidth]{images/ma_grader.pdf}%
%     \caption{A schematic diagram of proposed models for automated speaking assessment.}
% \label{fig:model_arch}
% \end{figure*}

\begin{figure*}[htbp]
    \centering
    \subfloat[]{
        \includegraphics[width=0.17\linewidth]{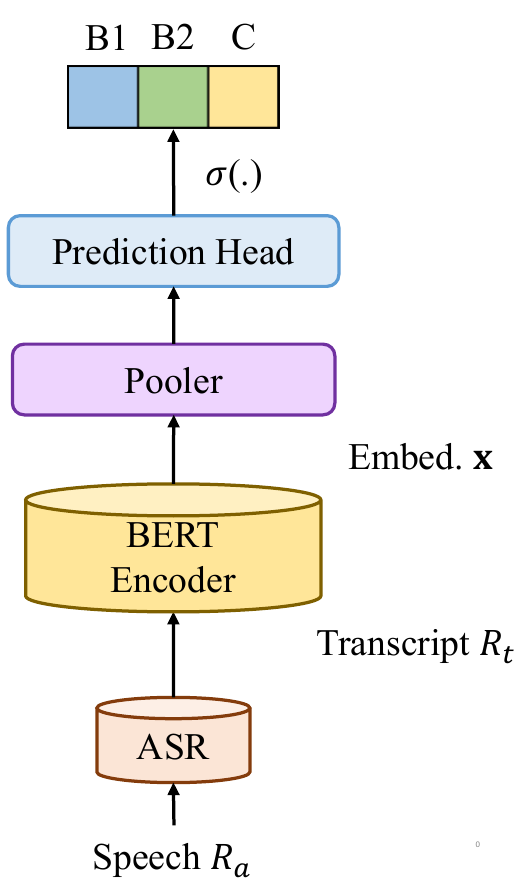}%
    \label{fig:bert_grader}}
    \hfil
    \subfloat[]{
        \includegraphics[width=0.17\linewidth]{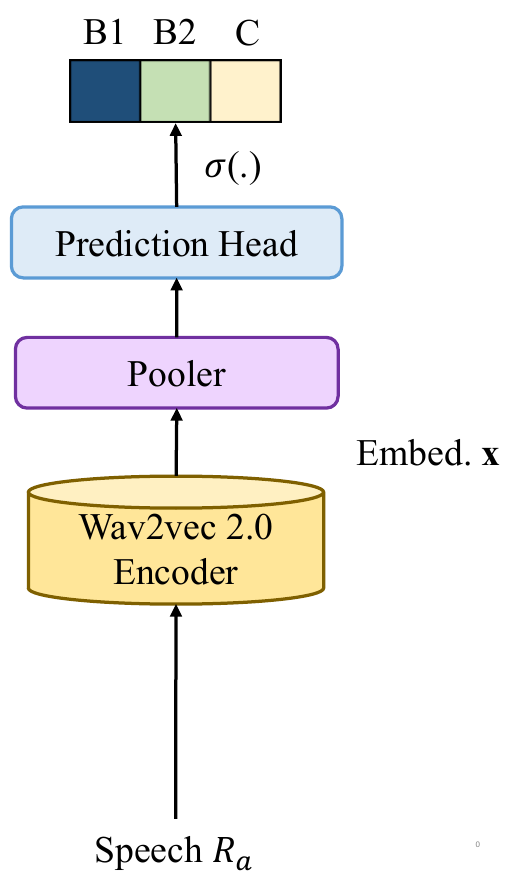}%
    \label{fig:wav2vec_grader}}
    \hfil
    \subfloat[]{
        \includegraphics[width=0.63\linewidth]{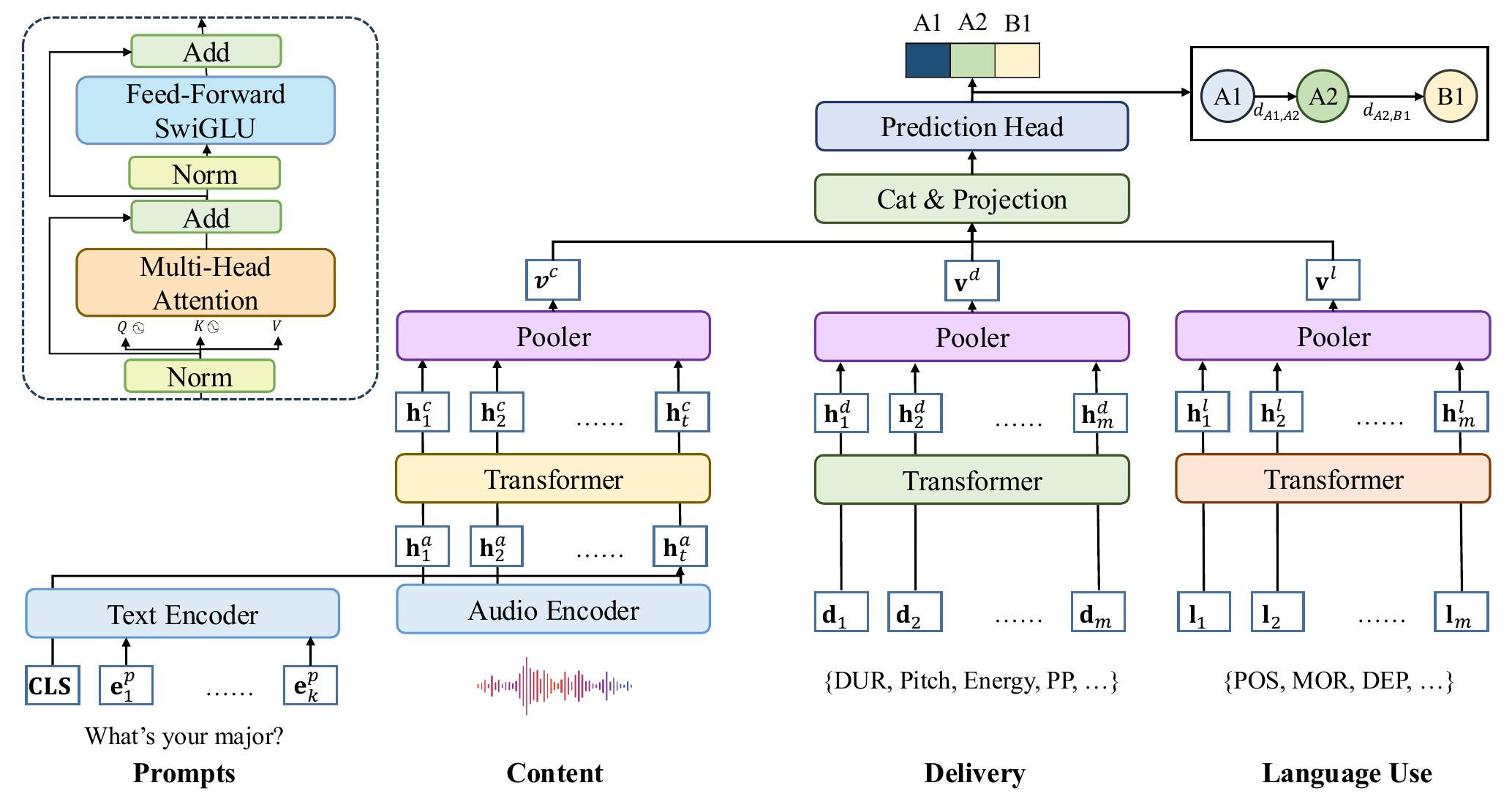}%
    \label{fig:ma_grader}}
    \caption{An overview of the model architecture for automated speaking assessment is presented. (a) depicts a text-based grader that utilizes ASR-generated transcripts and BERT embeddings to perform CEFR-level classification. (b) illustrates a speech-based grader that directly encodes raw audio input using wav2vec 2.0. (c) presents our multi-aspect framework, which models content, delivery, and language use with separate Transformer encoders. The resulting representations ($\mathbf{v}^*$) are concatenated, pooled, and passed to a prediction head  trained with both cross-entropy and MMO loss.}
\label{fig:model_arch}
\end{figure*}

\section{Methodology}

In this section, we formulate automated speaking assessment (ASA) as a classification task with respect to the spoken responses of L2 learners. Each training, validation or test instance is represented as an input-output pair $(\mathbf{x}_i, y_i)$, where $\mathbf{x}_i$ denotes the input features derived from multiple modalities of a learner’s response, including the raw audio signal $\mathbf{a}$, the ASR-generated transcription $\mathbf{w}$, and the given prompt $\mathbf{p}$, and $y_i$ the corresponding (or predicted) CEFR label. Each spoken response is recorded as a raw audio sequence $\mathbf{a} = \{a_1, a_2, \dots, a_t\}$, where $t$ denotes the number of acoustic frames. To extract semantic content, an ASR system is applied to generate a word-level transcription $\mathbf{w} = \{w_1, w_2, \dots, w_m\}$, with $m$ representing the number of recognized words. Additionally, each response is associated with a prompt $\mathbf{p} = \{p_1, p_2, \dots, p_k\}$, which provides contextual information (e.g., question) for the assessment task. Prior SSL-based ASA approaches typically rely on either the speech modality $\mathbf{a}$ (\textit{cf.} Figure \ref{fig:wav2vec_grader}) or the textual modality $\mathbf{w}$ (\textit{cf.} Figure~\ref{fig:bert_grader}) for proficiency prediction. The corresponding (or predicted) label $y_i \in \mathcal{Y} = \{1.0, 1.5, 2.0, \dots, 5.0\}$ represents a CEFR-aligned proficiency score, which is subsequently digitized to $\{1, 2, \dots, 8\}$ for classification over an 8-level scale spanning Pre-A1 to B2.

\subsection{Multi-aspect Proficiency Modeling}

As shown in Figure~\ref{fig:ma_grader}, the proposed architecture uses separate Transformer~\cite{NnVaswani2017_nips} encoders with minor adaptations \cite{NnLlama3rAttafiori2024_arxiv} to individually model the aspects of content, delivery, and language use, allowing the generation of aspect-specific representations that will be effectively integrated for scoring.

\textbf{\textit{Content Module}}: The content module aims to capture the semantic alignment between a learner's spoken response and the given prompt. The prompt $\mathbf{p}$ is encoded using a pre-trained BERT model to derive a sentence-level embedding from the [CLS] token. Simultaneously, frame-level speech features are extracted from the audio raw waveform $\mathbf{a}$ using wav2vec 2.0 (W2V), resulting in a sequence of contextual representations:
\begin{equation}
\mathbf{e}^\text{[CLS]} = \mathrm{BERT}([[CLS];\mathbf{p}]),
\end{equation}

\begin{equation}
\mathbf{h}^a_{1:t} = \mathrm{W2V}(\mathbf{a}).
\end{equation}
The prompt embedding $\mathbf{e}^\text{[CLS]}$ is replicated and concatenated with each frame-wise speech feature is fed into the Transformer encoder layer:
\begin{equation}
\mathbf{h}^c_{1:t} = \mathrm{Transformer}_\text{content}([\mathbf{e}^\text{[CLS]}; \mathbf{h}^a_1], \dots, [\mathbf{e}^\text{[CLS]}; \mathbf{h}^a_t]),
\end{equation}

\begin{equation}
\mathbf{v}^c = \mathrm{Pooler}(\mathbf{h}^c_{1:t}), 
\end{equation}

where attention pooling $\mathrm{Pooler}$ is applied over $\mathbf{h}^c_{i:t}$ to obtain the content representation $\mathbf{v}^c$, which encapsulates the content-related relevance between the spoken response and the prompt.

\textbf{\textit{Delivery Encoder}}: The delivery encoder captures temporal characteristics of spoken delivery, focusing on prosodic variation and phonetic clarity. Each word-aligned segment from ASR output is represented by a delivery feature vector $\mathbf{d}_i \in \mathbb{R}^{16}$, which includes pitch and energy statistics (mean, standard deviation, median, median absolute deviation, sum, maximum, minimum), segment duration, and confidence scores. The delivery vector sequence ${\mathbf{d}_{1:m}}$ is processed by a Transformer layer to obtain contextualized embeddings. An attention-based pooling layer is then applied to produce a fixed-dimensional vector $\mathbf{v}^d$ representing delivery-related information:
\begin{equation}
\mathbf{v}^d = \mathrm{Pooler}(\mathrm{Transformer}_\text{delivery}(\mathbf{d}_{1:m})). 
\end{equation}

\textbf{\textit{Language Use Module}}: This module is designed to capture lexical choices and syntactic structures present in the spoken response of the learner. Given the transcript produced by the ASR system, word-level linguistic annotations are obtained using the Stanza NLP toolkit \cite{ToolStanzaZhang2021_jamai}, including part-of-speech (POS) tags, dependency relations (DEP), and morphological features (MOR). These features are encoded into a sequence of linguistic feature vectors $\mathbf{l}_{1:m} \in \mathbb{R}^{m \times 263}$, which is then passed through a Transformer encoder to capture contextual dependencies across words. An attention-based pooling mechanism is applied to the contextualized sequence to produce a fixed-length embedding representing language use:
\begin{equation}
\mathbf{v}^l = \mathrm{Pooler}(\mathrm{Transformer}_\text{lang}(\mathbf{l}_{1:m})).
\end{equation}

To construct a holistic representation of the spoken response, the three aspect-specific embeddings, i.e., $\mathbf{v}^c$ (content), $\mathbf{v}^d$ (delivery), and $\mathbf{v}^l$ (language use), are concatenated and passed through a linear projection layer. The fused vector is subsequently fed into a prediction head to compute the logits over CEFR proficiency levels, denoted as $\mathbf{z} \in \mathbb{R}^C$:

\begin{equation}
\mathbf{z} = \mathrm{PredictionHead}(\mathrm{Projection}([\mathbf{v}^c; \mathbf{v}^d; \mathbf{v}^l])). 
\end{equation}

\subsection{Multi-Margin Ordinal Loss}
\label{subsec:mmol_loss}
To account for both the ordinal structure and non-uniform intervals inherent in CEFR-based scoring, this work introduces a logit-based \textit{multi-margin ordinal (MMO)} loss. While a previous study  \cite{LossClocPitawela2025_cvpr} has implemented multi-margin constraints at the hidden representation level to model ordinality for image classification, the proposed approach differs from it by imposing these constraints directly at the logit level. This design choice facilitates more explicit supervision of prediction outputs and offers improved alignment with the asymmetric progression of CEFR levels. To our knowledge, we are the first to extend and conceptualize this notion for use in ASA.

For each input instance, the MMO loss is defined as a pairwise constraint applied over sets of positive $(\mathbf{z}, \mathbf{z}_j) \in \mathcal{S}^+$ and negative $(\mathbf{z}, \mathbf{z}_k) \in \mathcal{S}^-$ logit pairs:
\begin{equation}
\mathcal{L}_{\mathrm{MMO}}(\mathbf{z}, y) = \max\left(0,\, d_{y,y_k} + \phi(\mathbf{z}, \mathbf{z}_k) - \phi(\mathbf{z}, \mathbf{z}_j)\right),
\end{equation}
\begin{equation}
\label{eq:dist}
d_{y,y_k} = d_{y,y+1} + \dots + d_{y_{k-1}, y_k},
\end{equation}
where $\phi(\cdot)$ denotes the cosine similarity function. The cumulative margin $d_{y,y_k}$ represents the ordinal distance between the ground-truth label $y$ and the negative label $y_k$, thereby enforcing greater separation in logit space for pairs of labels that are more distant on the CEFR scale.

The final loss function integrates the MMO loss with the conventional cross-entropy objective to jointly optimize classification accuracy and ordinal consistency:
\begin{equation}
\mathcal{L} = \lambda \cdot \mathcal{L}_{\mathrm{CE}} + (1 - \lambda) \cdot \mathcal{L}_{\mathrm{MMO}},
\end{equation}
where the hyperparameter $\lambda \in [0,1]$ controls the balance between standard classification supervision and ordinal-aware learning.

\section{Experimental Setup}
\subsection{Implementation Details}
Model configurations were initialized using pretrained models from the HuggingFace Transformers library \cite{ToolHugefaceWolf2020_emnlp}. Two SSL-based models, \texttt{bert-base-uncased}\footnote{\url{https://huggingface.co/bert-base-uncased}} and \texttt{wav2vec2-base}\footnote{\url{https://huggingface.co/facebook/wav2vec2-base}}, were employed as text and speech encoders, respectively. For all Transformer encoders, the number of attention heads was set to 1 to encourage lightweight modeling and reduce overfitting.  The weighting coefficient $\lambda$ in the objective function was tuned via grid search on the validation set, with $\lambda = 0.5$ selected based on empirical performance across evaluation metrics. All models were trained on an NVIDIA 3090 GPU using AdamW optimizer, with a batch size of 32 and an initial learning rate of 1e-4. The training process of the all classifier was stopped early with 30 patience epochs based on the averaged macro-averaged score from the validation set.

\subsection{Evaluation Metrics}
Tangible evaluations of the effectiveness of ASA models are crucial for grading applications, for which accurate prediction at all levels is essential. However, as the distribution of CEFR levels is unbalanced, conventional evaluation metrics such as accuracy (ACC) may underestimate the performance of ASA models. Therefore, macro-averaged F1 score is used to penalize those models that treat the minor classes poorly.

\begin{table}[!t]
\centering
\caption{Model performance on the TEEMI test set.}
\label{tab:overall_performance}
\resizebox{\columnwidth}{!}{%
\begin{tabular}{lcccccccc}
\hline
\multirow{2}{*}{\textbf{Models}} &
  \multicolumn{2}{c}{\textbf{Content}} &
  \multicolumn{2}{c}{\textbf{Delivery}} &
  \multicolumn{2}{c}{\textbf{Language use}} &
  \multicolumn{2}{c}{\textbf{Holistic}} \\ \cline{2-9} 
                                 & \textbf{ACC↑}  & \textbf{F1↑}   & \textbf{ACC↑}  & \textbf{F1↑}   & \textbf{ACC↑}  & \textbf{F1↑}   & \textbf{ACC↑}  & \textbf{F1↑}   \\ \hline
W2V~\cite{AsaBanno2022_slt}      & 35.08          & 29.55          & 39.92          & 37.11          & 36.29          & 31.53          & 34.67          & 30.17          \\
BERT~\cite{AsaBanno2022_slt}     & 33.47          & 28.31          & 37.90          & 31.19          & 36.29          & 31.66          & 35.48          & 31.19          \\
W2V-BERT~\cite{AsaPark2023_asru} & 35.08          & 27.83          & 38.31          & 31.46          & 41.13          & 35.15          & \textbf{38.71} & 30.35          \\
W2V-PT~\cite{AsaLo2024_naacl}    & 30.24          & 24.23          & 38.71          & 34.33          & \textbf{42.74} & 36.00          & 34.68          & 29.87          \\
BERT-PT~\cite{AsaLo2024_naacl}   & 29.44          & 27.25          & 40.73          & 37.22          & 35.08          & 33.90          & 33.87          & 32.49          \\ \hline
MA                               & 35.89          & 31.60          & 41.53          & 39.04          & 38.31          & 31.83          & 33.87          & 26.28          \\
MA + MMO                         & \textbf{37.10} & \textbf{34.77} & \textbf{42.34} & \textbf{40.87} & 42.34          & \textbf{40.22} & 36.29          & \textbf{35.55} \\ \hline
\end{tabular}%
}
\end{table}

\begin{table}[!t]
\centering
\caption{Model performance on the unseen test dataset.}
\label{tab:unseen_performance}
\resizebox{\columnwidth}{!}{%
\begin{tabular}{cccccccccc}
\cline{2-10}
\multirow{2}{*}{\textbf{Models}} &
  \multirow{2}{*}{\textbf{Task}} &
  \multicolumn{2}{c}{\textbf{Content}} &
  \multicolumn{2}{c}{\textbf{Delivery}} &
  \multicolumn{2}{c}{\textbf{Language use}} &
  \multicolumn{2}{c}{\textbf{Holistic}} \\ \cline{3-10} 
 &
   &
  \textbf{ACC↑} &
  \textbf{F1↑} &
  \textbf{ACC↑} &
  \textbf{F1↑} &
  \textbf{ACC↑} &
  \textbf{F1↑} &
  \textbf{ACC↑} &
  \textbf{F1↑} \\ \hline
\multirow{3}{*}{MA}   & A02 & 30.08 & 31.91 & 39.84 & 39.12 & 34.15 & 36.02 & \textbf{32.52} & \textbf{35.30} \\
                      & B02 & \textbf{32.66} & 27.76 & 36.29 & 34.30 & \textbf{34.27} & 28.89 & \textbf{33.47} & 27.63 \\
                      & C01 & 20.17 & 14.30 & 26.05 & 21.85 & 37.82 & 34.61 & 25.21 & 21.27 \\ \hline
\multirow{3}{*}{+MMO} & A02 & \textbf{35.77} & \textbf{35.77} & \textbf{43.09} & \textbf{44.84} & \textbf{39.02} & \textbf{40.44} & 30.89 & 32.32 \\
                      & B02 & 31.85 & \textbf{29.54} & \textbf{38.71} & \textbf{37.74} & 30.65 & \textbf{31.24} & 32.66 & \textbf{31.40} \\
                      & C01 & \textbf{30.25} & \textbf{29.62} & \textbf{44.54} & \textbf{43.04} & \textbf{39.50} & \textbf{35.27} & \textbf{31.93} & \textbf{31.14} \\ \hline
\end{tabular}%
}
\end{table}

\section{Results and Discussion}

\subsection{Overall Performance}
At the outset, we report on the performance of baseline systems and proposed models on the TEEMI corpus, evaluated across four CEFR-based scoring aspects: content, delivery, language use, and holistic proficiency. The baseline models compared here are speech-based SSL (W2V) \cite{AsaBanno2022_slt}, text-based SSL (BERT) \cite{AsaBanno2022_slt}, a multimodal fusion of both (W2V-BERT) \cite{AsaPark2023_asru}, and their respective prototypical variants (W2V-PT and BERT-PT) \cite{AsaLo2024_naacl}. As shown in Table~\ref{tab:overall_performance}, W2V-BERT demonstrates the strongest performance in terms of holistic proficiency, achieving the highest absolute accuracy (38.71\%), thereby highlighting the benefit of integrating acoustic and linguistic cues. While prototypical models show competitive F1 scores on the assessment of the aspects of delivery and language use, their performance degrades substantially on the aspects of content and holistic proficiency, indicating limited generalizability across aspects. The proposed multi-aspect (MA) framework, which integrates handcrafted features with SSL-derived embeddings, consistently outperforms all baseline systems, with respect to macro-averaged F1, across the four scoring aspects. Particularly strong performance is observed on the delivery aspect, alongside competitive accuracy on the aspects of content and language use. Further improvements emerge with the introduction of the Multi-Margin Ordinal (MMO) loss. The MA+MMO variant achieves the highest holistic F1 score (35.55\%) and yields absolute gains of 3.17 and 8.39 in macro-averaged F1 for content and language use, respectively, compared to the base MA model. This significant boosts of performance confirms the promising potential of our proposed modeling strategies for ASA.

\begin{figure*}[!t]
    \subfloat[Content (MA)]{
        \includegraphics[width=0.23\linewidth]{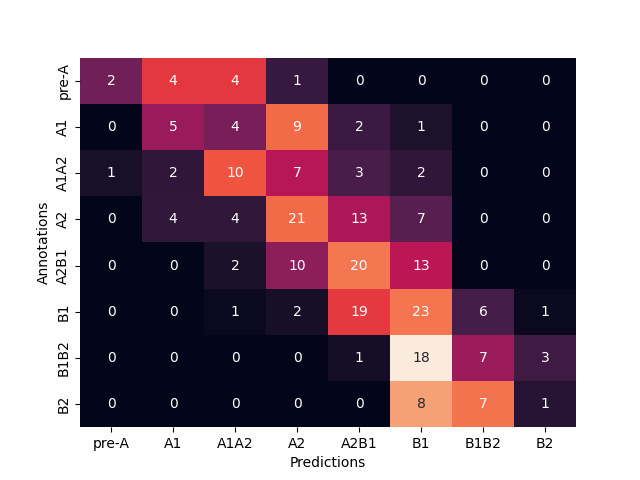}%
    \label{fig:cm_ma_asa_content}}
    \hfil
    \subfloat[Delivery (MA)]{
        \includegraphics[width=0.23\linewidth]{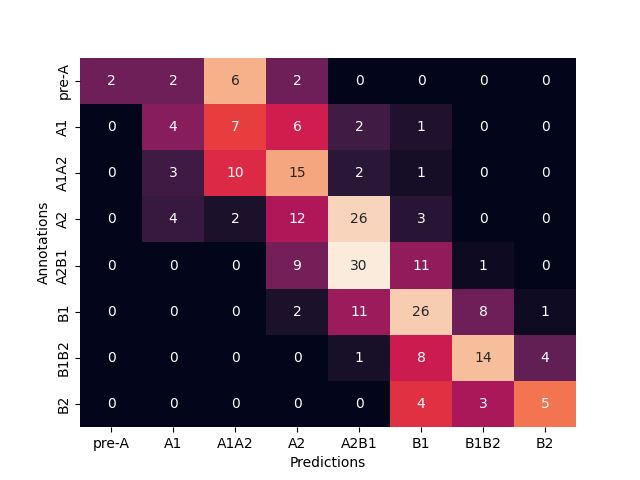}%
    \label{fig:cm_ma_asa_pronunciation}}
    \hfil
    \subfloat[Language Use (MA)]{
        \includegraphics[width=0.23\linewidth]{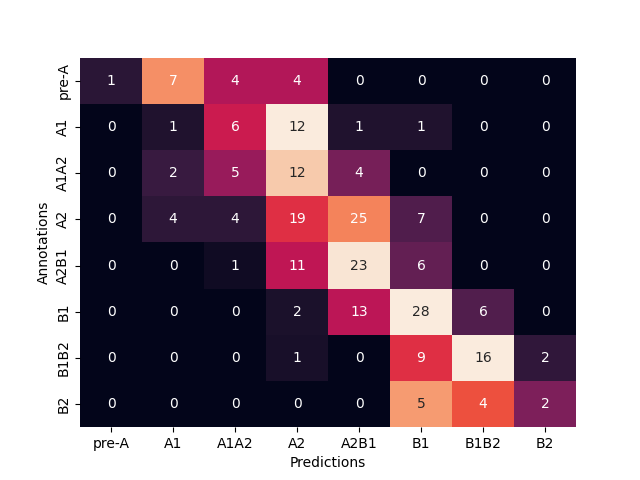}%
    \label{fig:cm_ma_asa_vocabulary}}
    \hfil
    \subfloat[Holistic (MA)]{
        \includegraphics[width=0.23\linewidth]{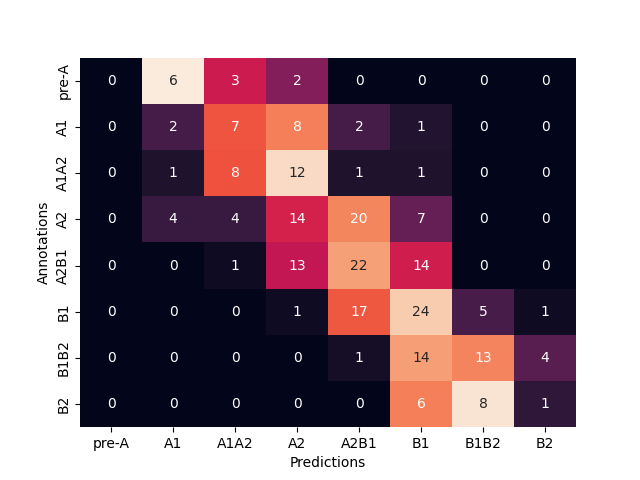}%
    \label{fig:cm_ma_asa_holistic}}
    \vfil
    \subfloat[Content (MA+MMO)]{
        \includegraphics[width=0.23\linewidth]{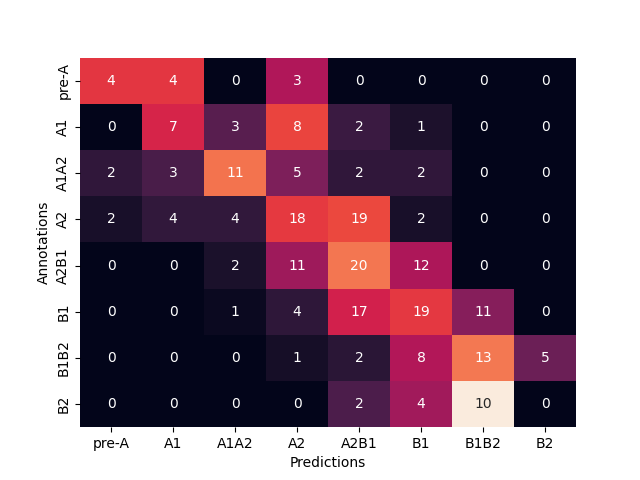}%
    \label{fig:cm_ma_asa_ocmm_content}}
    \hfil
    \subfloat[Delivery (MA+MMO)]{
        \includegraphics[width=0.23\linewidth]{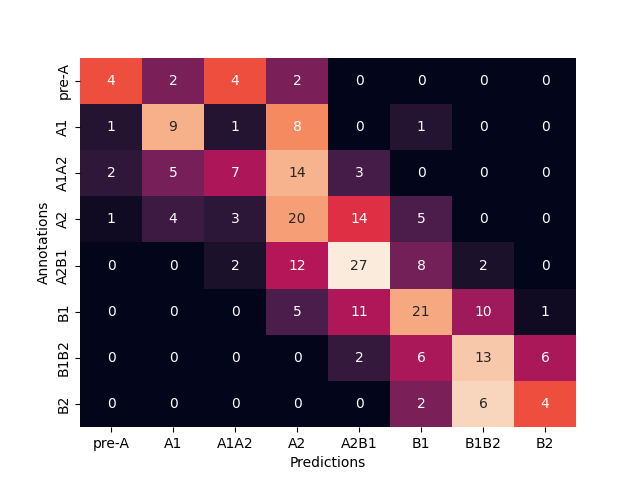}%
    \label{fig:cm_ma_asa_ocmm_pronunciation}}
    \hfil
    \subfloat[Language Use (MA+MMO)]{
        \includegraphics[width=0.23\linewidth]{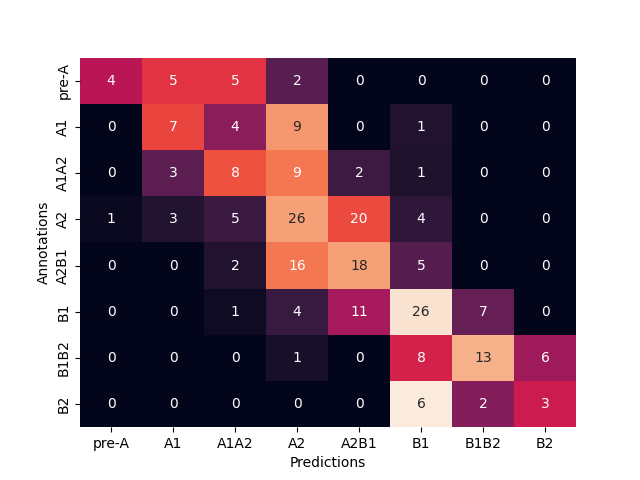}%
    \label{fig:cm_ma_asa_ocmm_vocabulary}}
    \hfil
    \subfloat[Holistic (MA+MMO)]{
        \includegraphics[width=0.23\linewidth]{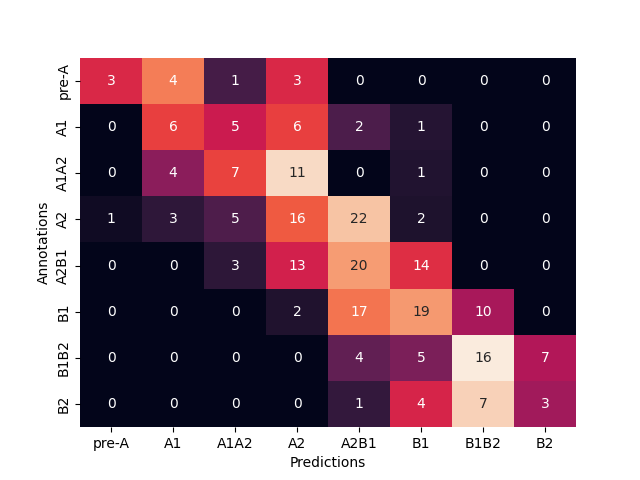}%
    \label{fig:cm_ma_asa_ocmm_holistic}}
    \caption{Confusion matrices comparing the performance of multi-aspect ASA classifiers with and without the proposed multi-margin ordinal (MMO) loss across four scoring aspects: content (a,e), delivery (b,f), language use (c,g), and holistic proficiency (d,h). The upper row (a–d) shows results from the MA model, and the lower row (e–h) from MA+MMO. Each matrix plots true CEFR levels (rows) against predicted levels (columns), illustrating both accuracy and ordinal alignment.}
\label{fig:cm_ma_asa}
\end{figure*}

\begin{figure}[htbp]
    \centering
    \includegraphics[width=0.80\linewidth]{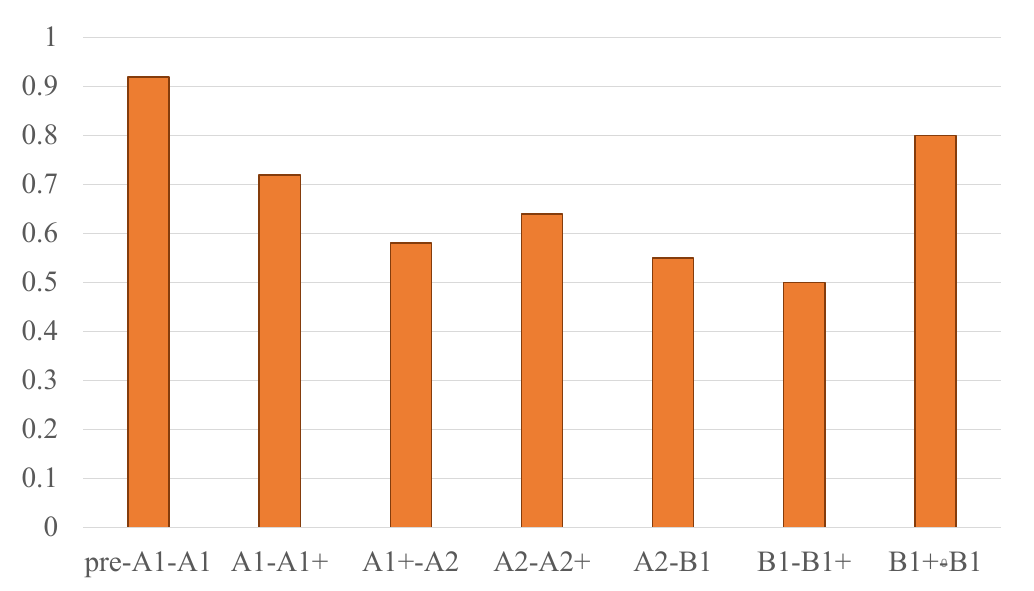}
    \caption{Pairwise distances between adjacent CEFR levels computed from training instances.}
    \label{fig:cefr_interval}
\end{figure}

\subsection{Evaluation on Unseen Prompts and Tasks}
To evaluate the generalizability of the proposed models, we conduct a second set of experiments on three unseen tasks from the TEEMI corpus: A02, B02, and C01, as outlined in Table~\ref{tab:teemi_dataset_split}. A02 shares the same task type as the training task A01 but differs in prompt content, whereas B02 and C01 involve both prompt and task-type shifts, corresponding to situational and thematic Q\&A tasks, respectively.

As illustrated in Table~\ref{tab:unseen_performance}, the MA+MMO model consistently outperforms the MA model across all CEFR levels. For A02, where variation arises primarily from the prompt, MA+MMO improves the overall F1 score by 1.02\% and the F1 score for language use by 4.42\%, demonstrating robustness to prompt-level differences. For B02 and C01, which introduce broader task variation, performance gains are more pronounced. MA+MMO achieves a 3.77\% increase in overall F1 score on B02, and on C01, substantial improvements are observed in delivery (+21.19\%) and overall F1 (+9.87\%). These findings indicate that the incorporation of the MMO loss enhances the model’s capacity to generalize to unseen prompts and task configurations.

\subsection{Visualization of Confusion Matrices}
\label{sec:confusion_matrix}

Figure~\ref{fig:cm_ma_asa} presents the confusion matrices for each scoring aspect under two configurations: the proposed multi-aspect (MA) model (top row) and its variant enhanced with the Multi-Margin Ordinal (MMO) loss (bottom row). These confusion matrices reveal the correspondence between predicted CEFR levels and reference labels, offering a detailed view of class-wise behavior and ordinal consistency.

The results obtained from the MA model reveal that frequent confusions occurs between adjacent levels, such as A2 versus A2+ and B1 versus B1+, particularly for the aspects of content and language use. An overestimate bias toward mid-range levels (e.g., A2+, B1) is also observed, likely due to imbalanced label distribution in the training set. The addition of the MMO loss results in clearer diagonal concentration and a notable reduction in errors between distant levels. These results indicate that the MMO loss introduces multiple margins that account for inter-level distances, thereby enhancing performance of the model prediction.

\subsection{Distance Analysis of CEFR Levels}

To further investigate the ordinal structure and latent geometry of CEFR proficiency levels, an analysis was conducted on the semantic distances between adjacent class labels. As depicted in Figure~\ref{fig:cefr_interval}, we compute the pairwise distances between CEFR levels based on Equation \ref{eq:dist}. The observed distances suggest marked non-uniformity across the CEFR scale. For instance, the semantic gap between Pre-A1 and A1 is the largest (0.9236), while the transition from B1 to B1+ is the smallest (0.4868). These findings empirically validate prior observations \cite{CefrNuiHheilman2008_bea}, confirming the claim that proficiency levels are not equidistant in label embedding space. The presence of such asymmetric transitions challenges the assumptions made in traditional ordinal classification methods that rely on fixed margins, underscoring the necessity of adopting flexible, data-driven strategies, such as the proposed MMO loss, to capture the non-uniform progression inherent in ASA.

\section{Conclusion and Future Work}

This paper has proposed two innovative modeling strategies for automated speaking assessment (ASA): multi-aspect classification and multi-margin ordinal (MMO) loss. The first strategy is designed to mitigate the modality constraints of SSL models when applied to ASA. The second strategy is compatible with multi-aspect modeling strategy while addressing the challenges of score ordinality and non-uniform level intervals. Experiments on the TEEMI dataset have demonstrated the effectiveness of our methods in relation to previous methods. Furthermore, evaluations on different unseen prompts confirm the generalizability of our model across different ASA tasks. For future work, we will plan to explore joint training strategies across prompts and scoring aspects to enhance ASA robustness on diverse tasks. In addition, we envisage integrating multi-modal large language models (MLLMs) that fuse acoustic and textual information, thereby advancing the comprehensiveness and interpretability of ASA.

\bibliographystyle{IEEEtran}
\bibliography{mybib}

\vspace{12pt}

\end{document}